\documentclass[twocolumn,preprintnumbers,amsmath,amssymb]{revtex4-1}
\usepackage{graphicx}
\usepackage{dcolumn}
\usepackage{bm}
\usepackage{hyperref}

\begin{document}

\title{Amorphous Precursors of Crystallization during Spinodal Decomposition}

\author{Leopoldo R. G\'omez$^{1,2}$}
\email{gomez@lorentz.leidenuniv.nl}
\author{Daniel A. Vega$^{1}$}

\affiliation{ $^1$Department of Physics and Instituto de F\'isica
del Sur, IFISUR (UNS-CONICET). Al\'em 1253. (8000) Bah\'ia Blanca,
Argentina.\\
$^2$Instituut-Lorentz, Universiteit Leiden, P. O. Box 9506, 2300
RA Leiden, The Netherlands.}

\date{\today}

\begin{abstract}
A general Landau's free energy functional is used to study the
dynamics of crystallization during liquid-solid Spinodal
Decomposition (SD). The strong length scale selectivity imposed
during the early stage of SD induces the appearance of small
precursors for crystallization with icosahedral order. These
precursors grow in densely packed clusters of tetrahedra through
the addition of new particles. As the average size of the
amorphous nuclei becomes large enough to reduce geometric
frustration, crystalline particles with a body center cubic
symmetry (bcc) heterogeneously nucleates on the growing clusters.
The volume fraction of the crystalline phase is strongly dependant
on the depth of quench. At deep quenches, the SD mechanism
produces amorphous structures arranged in dense polytetrahedral
aggregates.

\end{abstract}

\maketitle

According to the classical theory of symmetry breaking phase
transitions, the crystallization process may proceed via two
distinct mechanisms, nucleation and growth (NG) or spinodal
decomposition (SD) \cite{Debenedetti}. In the NG mechanism the
initial disordered phase is metastable and the relaxation of the
system is promoted by the overcoming of a free energy barrier.
This process involves the formation of a critical nucleus of the
crystalline phase by structural fluctuations. Differently from the
NG process, SD does not require large fluctuations to initiate the
phase transition and it is characterized by the exponential growth
of density fluctuations of a dominating wavelength, entirely
determined by the thermodynamic properties of the system. At early
times this process leads to the formation of disordered isotropic
states that eventually evolve towards the equilibrium phase
through coarsening \cite{Vega}, \cite{Gomezprl2006}.

Although this classical picture provides a qualitative description
of the phenomenology involved in the phase transition process,
during the last thirty years, experiments, theory and simulations
have shown that the process of crystallization is far from being
trivial. For example, in the NG regime, the kinetic pathway
towards equilibrium may involve intermediate phases with a
symmetry different than the corresponding to equilibrium
\cite{WeitzScience}-\cite{Lutsko}. According to the Ostwald's step
rule the nucleated phase is not necessarily the thermodynamically
most stable, but the energetically closest to the disordered state
\cite{FrenkelPRL}. A similar process was observed in the
crystallization of globular proteins and colloids, where density
inhomogeneities in the fluid lower the free energy barrier for NG
\cite{FrenkelScience}-\cite{Lutsko}. On the other hand, the
relaxation in the spinodal region is still poorly understood. For
example, it has been recently observed that thermal fluctuations
play an important role in the dynamics of SD. In the neighborhood
of the spinodal line it was found that the equilibrium phase can
be pseudo-nucleated by density wave fluctuations \cite{VegaGomez}.
A complex phase separation kinetics have been observed also in
different systems, like colloids and polymers, where it has been
found that a spinodal-like dynamics can precedes the
crystallization process \cite{VegaGomez}, \cite{PickeringSnook}.

In this work a continuous Landau's free energy expansion and a
relaxational dynamics are used to study the kinetics of
crystallization of a bcc structure in the spinodal region. As
compared with molecular dynamics simulations, the phase field
approach employed here naturally incorporates the elasticity of
the bcc crystals and provides an efficient approach over diffusive
time scales, while the identification of the precursors of
crystallization as well as amorphous and crystalline regions can
be easily determined through the local amplitude of the continuum
order parameter.

Symmetry-breaking phase transition in a wide variety of systems
can be studied through an expansion of the free energy $\Phi$ in
terms of an appropriated order parameter $\psi$. Here we employ
the expansion \cite{Andelman}:
\begin{equation}
\Phi=\int \{W(\psi)+D(\nabla \psi)^{2}+ b \int
\frac{\psi(\textbf{r})\,\psi(\textbf{r}')}{|\textbf{r}-\textbf{r}'|}dV'\}\,dV,
\end{equation}
where $W(\psi)=-\tau \psi^{2}+\nu \psi^{3}+\lambda \psi^{4}$. Here
the order parameter represents the deviation of the particle
density from the uniform value characteristic of the liquid state
(where $\psi=0$). The parameter $\tau$ is proportional to $T_S -
T$, being a measurement of the deep of quench, with $T_S$ the
spinodal temperature where the continuous phase transition begins.
The constants $\nu$ and $\lambda$ are related to the symmetry and
saturation of $\psi$ at equilibrium, and $D$ is a free energy
penalization to spatial variations of the order parameter
\cite{Vega}, \cite{Gomezprl2006}, \cite{Shi}.

The dynamics of SD can be studied through a Cahn-Hilliard equation
\cite{CahnHilliard}: $\partial \psi / \partial
t=M\nabla^{2}\{\delta \Phi / \delta \psi\}$, where $M$ is a
mobility coefficient \cite{Supp}.

\begin{center}
\begin{figure}[t]
\includegraphics[width=8.5 cm]{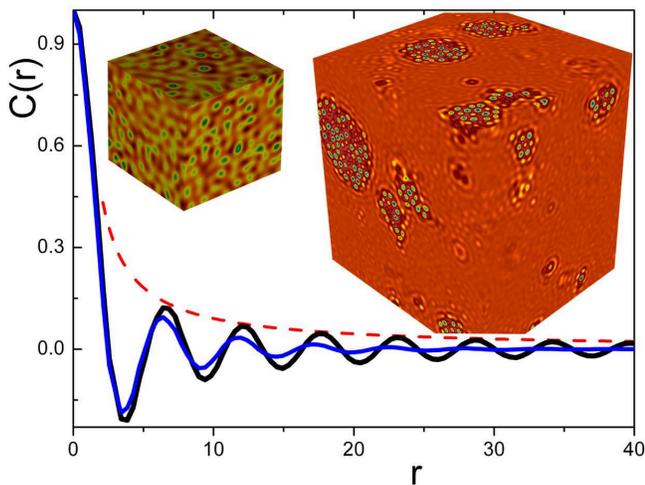} \caption{
 Early autocorrelation function $C(r)$ for a system quenched into the
 spinodal region with reduced temperatures $\tau_r=(\tau-\tau_S)/\tau_S$ $=1$x$10^{-5}$ (black line) and $\tau_r=1$x$10^{-1}$ (blue line).
The dashed red line indicates $C(r) \sim 1/r$. The inset shows the
early time real space fluctuations (left) and the formation of
precursor by the non-linear enhancement of density fluctuations
(right).}
\end{figure}
\end{center}

Since the initial disordered state is characterized by small
fluctuations ($\psi \sim 0$), the early dynamics is almost linear
and the system's state can be described as a random superposition
of density waves of the form \cite{VegaGomez}:
$\psi(\textbf{r},t)= \sum_{\textbf{k}} A_{\textbf{k}} exp \,{(i \,
\textbf{k}\cdot \textbf{r}+\lambda (k) \, t)}$. Here
$A_{\textbf{k}}$ is the initial amplitude of the $\textbf{k}$-mode
and the amplification factor $\lambda(k)=- D\,k^4 + \tau\,k^2-b$
selects the range of unstable modes (those modes for which
$\lambda(k)>0$). In real space the system display a disordered
pattern characterized by dominating length scale related with the
most unstable modes \cite{VegaGomez}.

\begin{center}
\begin{figure}[b]
\includegraphics[width=8.5cm]{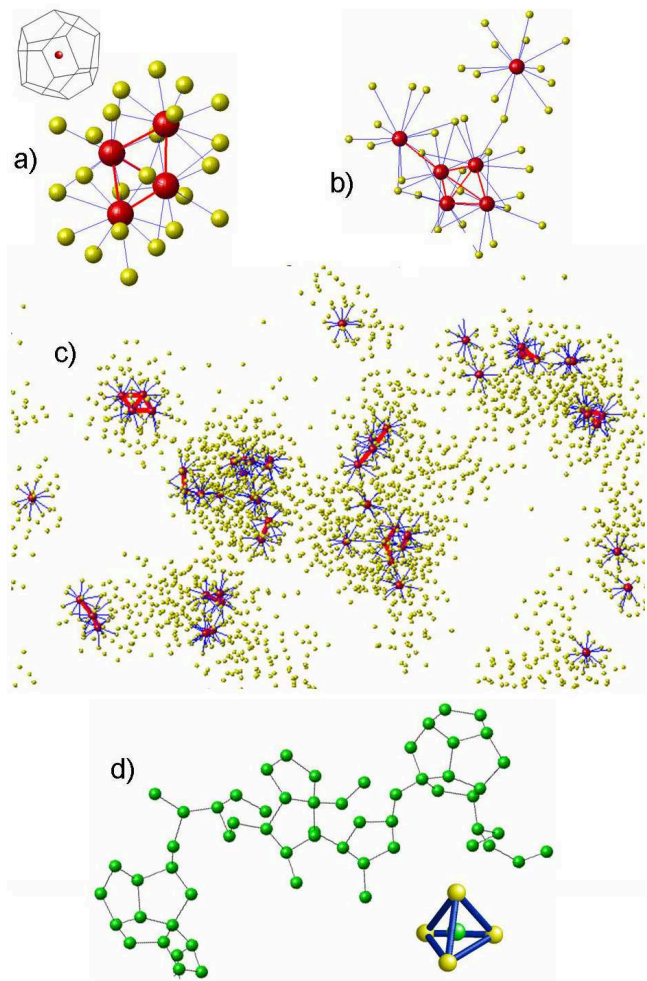} \caption{
(a) Precursor formed by four icosahedral particles (red) and their
first neighbors (yellow). Red lines indicate bonds between
icosahedral particles and blue lines between icosahedra and first
neighbors. Inset: Voronoi diagram of an icosahedral particle
(dodecahedron). (b) Precursor formed by a compact cluster of
icosahedra arranged in the vertex of a tetrahedron. (c)
Propagating amorphous nuclei (yellow) having compact clusters and
small chains of icosahedral particles (red). (d) Regular
tetrahedron (down right), where yellow spheres indicate the
particles and green sphere indicates the tetrahedron center's, and
polytetrahedral aggregate (up), where black lines connect
tetrahedra with common faces.}
\end{figure}
\end{center}

For systems quenched near below the spinodal line, where
$\tau_S=2\sqrt{bD}$, the region of unstable modes becomes sharply
peaked around a dominating wave vector amplitude $k_0 \sim
\sqrt{\tau/D}$. In this case the random superposition of modes
leads to the emergence of a strongly correlated filamentary
network of density wave fluctuations, with strong similarities to
those found in quantum billiards and other physical phenomena
involving random wave superposition \cite{VegaGomez}. Figure 1
shows the long range density correlations that emerge in the early
state. In this case, the azimuthally averaged two point
correlation function, $C(r)=\langle \int d\textbf{r'}
\psi(\textbf{r'}) \psi(\textbf{r+r'})\rangle$, of a critically
quenched system behaves like a Bessel function decaying as
$C(r)\sim 1/r$, in agreement with the theoretical predictions for
the random superposition of waves \cite{VegaGomez}. For
sub-critical systems ($\tau>\tau_{s}$) the network of fluctuations
loose correlation and $C(r)$ decays faster. As shown below, the
early correlations in $\psi$ have a profound effect in the later
evolution of the system.

As time proceeds, there is a continuous amplification of $\psi$
until the anharmonic terms of the free energy functional cannot be
neglected and nonlinear dynamics comes into play. While at deep
quenches there is a lack of correlation in $\psi$ and the
classical picture of SD is recovered, at shallow quenches the
early network of density wave fluctuations trigger the
inhomogeneous appearance of precursors for crystallization (Figure
1). Note that the local symmetry of the precursors is dictated by
non-linear dynamic effects and does not necessarily coincides with
the symmetry of the phase of equilibrium.

In order to track the evolution of the system we identify
structural features by means of Voronoi diagrams through the
centers of the particles (bulk particles are defined as local
maxima of $\psi$). Voronoi diagrams indicate that the early
precursors are constituted by compact clusters and chains of
icosahedrally arranged particles (Figs. 2a y 2b). That is, the
symmetry of the precursors is dictated by the non-linear dynamics
rather than by equilibrium states.

The presence of precursors with icosahedral symmetry in the early
stage of SD is not surprising. One of the oldest scenarios for the
formation of amorphous matter and glasses is geometrical
frustration. According to this scenario, the amorphization of
supercooled liquids is related with their tendency to form locally
compact structures, like icosahedra, incompatible with
translational symmetries \cite{Frank}-\cite{Royall}. The symmetry
found in the precursors is also in agreement with the
Alexander-McTague prediction which indicates that local
icosahedral order is favored in the absence of a NG mechanism
\cite{AlexanderMcTague}.

Once the precursors have been formed, they begin to grow with an
approximately spherical shape at a constant rate by the addition
of new particles (Fig. 2c). Given that the initial symmetry is
incompatible with translational order, the aggregation process
produces growing amorphous clusters without any appreciable
symmetry. As the  size of the precursors increases, the
geometrical frustration is reduced and a number of particles begin
to aggregate with a crystalline structure at the front of the
propagating precursor (Fig. 3a) \cite{Comment}.

\begin{center}
\begin{figure}[t]
\includegraphics[width=8.5cm]{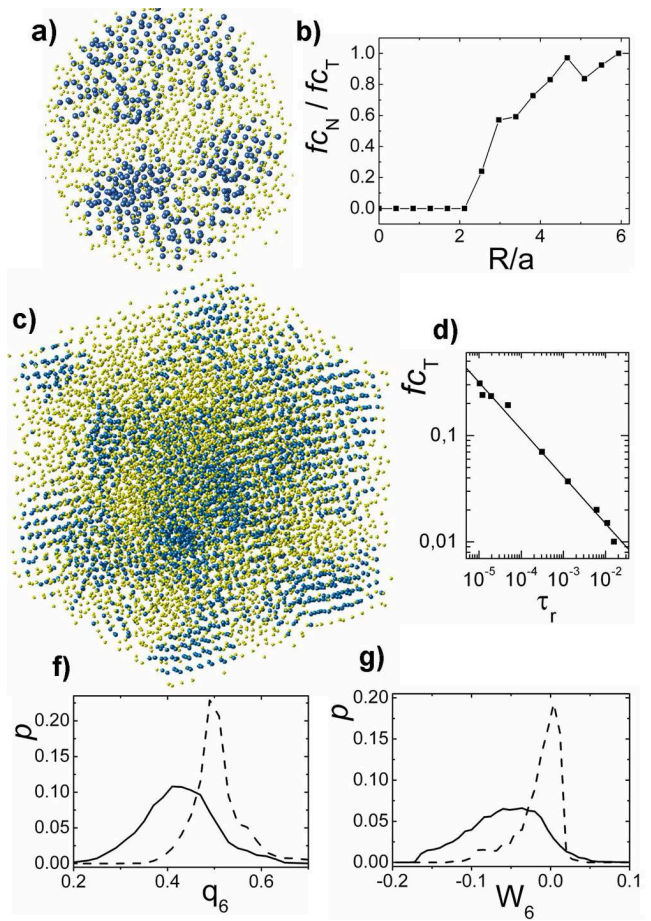}
\caption{ (a) Propagating nucleus formed by an amorphous core
(yellow particles) and crystalline particles (blue particles). (b)
Typical nuclei density profile. (c) System composed by amorphous
and crystalline regions after the growth and collision of the
different nuclei. (d) Fraction of crystalline particles as a
function of $\tau_r$. (f)-(g) $q_6$ and $W_6$ distributions for
the amorphous (continuous line) and crystalline (dotted line)
regions.}
\end{figure}
\end{center}

This process of surface crystalline nucleation shows similarities
with Monte Carlo results for the heterogeneous nucleation on the
surface of spherical colloidal seeds \cite{FrenkelNature}. In
agrement with the Monte Carlo simulations, here we also observe
that the crystalline particles do not span the whole surface of
the precursor because crystals cannot grow without generating
amorphous regions. Then, the growing nuclei shows an amorphous
core surrounded by a shell constituted by amorphous and
crystalline phases (Fig. 3a). A similar phenomena has been also
observed in the freezing gold nanoclusters where initially the
system crystallize on the surface of the nuclei while the
amorphous core with icosahedrally arranged particles remains
stable for some time \cite{Nam}.

Figure 3b shows the average radial profile of the ratio between
the fraction of crystalline particles on the propagating nuclei
$\emph{fc}_N$ and the final fraction of crystalline particles
throughout the system $\emph{fc}_T$. This figure shows that
crystal particles begin to nucleate at the surface of the
amorphous precursors for radius of the order of $r_c \sim (3-5)
a$, with $a$ the inter-particle average distance, also in good
agreement with the results of Cacciuto et al.
\cite{FrenkelNature}. Then for $r<r_c$ the high curvature of the
precursors clearly frustrates the formation of crystal bonds. The
subsequent growth and collision of the different propagating
nuclei leads to the formation of a structure having crystalline
and amorphous regions (Fig. 3c).

Once the SD process has been completed, the volume fraction of
crystals in the system depends strongly on temperature. As the
depth of quench increases, there is a larger number of unstable
modes in the system and $C(r)$ decays faster than $1/r$ (Fig. 1).
Consequently, as the temperature of quench drops, the number of
amorphous precursors increases, incrementing in this way the
fraction of amorphous material. Figure 3d shows the fraction of
crystalline particles $\emph{fc}_T$ as a function of the reduced
temperature $\tau_r=(\tau-\tau_s)/\tau_s$. We found that
$\emph{fc}_T$ follows a power law with $\tau_r$ ($\emph{fc}_T \sim
\tau_r^{-1/2}$). This behavior is consistent with previous results
indicating that the average distance between precursors $\xi$
diverge as $\xi \sim \tau_r^{-\eta}$ ($\eta \sim 1/5$,
$\emph{fc}_T \sim \xi^{3}$) \cite{VegaGomez}.

To identify the symmetry of the crystalline phase surrounding the
amorphous precursor we applied the bond order parameter analysis
introduced by Steinhardt \emph{et al} \cite{Steinhardt}, where the
local structure is characterized in terms of the symmetry of near
neighbor bonds by using spherical harmonics. The distributions of
the bond order parameters are very sensitive to the underlaying
symmetry, allowing a clear identification of the structure.
Figures 3f and 3g show the distributions of the rotational
invariants $q_6$ and $w_6$ , for the amorphous and crystalline
regions. These distributions reveal that crystal particles arrange
in a bcc lattice (peaks at $q_6 \sim 0.5$ and $w_6 \sim 10^{-3}$)
while in the amorphous phase both distributions become broader and
shifted towards the low bond order parameter region
\cite{WeitzScience}, \cite{Steinhardt}.

At long times the relaxation of the system continues through a
complex process involving structural changes (coarsening), where
there is an invasion of the crystalline structure on the amorphous
regions. However, at low temperatures the dynamics can be highly
arrested and the structure freezes.

\begin{center}
\begin{figure}[t]
\includegraphics[width=8.5cm]{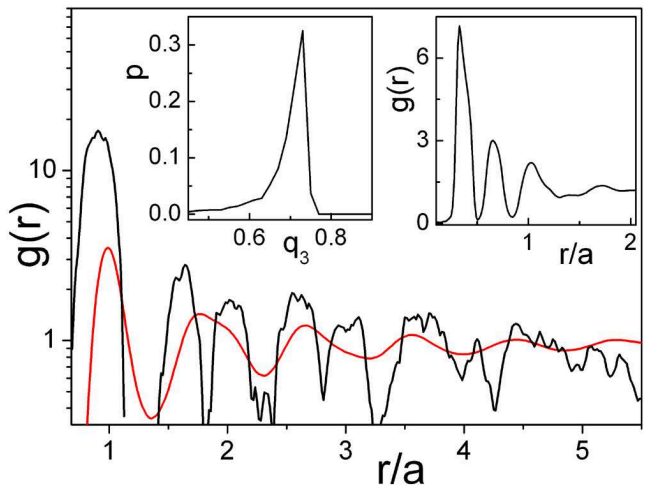}
\caption{Pair correlation functions $g(r)$ calculated with the
centres of all the particles (red) and with the centres of the
icosahedra (black). Inset: distribution for the $q_3$ bond order
parameter (left) and the pair correlation function calculated
through the center of the regular tetrahedra (right).}
\end{figure}
\end{center}

In order to study the  glasslike amorphous structure in the deep
spinodal region,  here we also analyze the structural features of
a system where the formation of crystals is totally frustrated
($\tau_{r} > 10^{-2} $ in Fig. 3). Figure 4 shows the pair
correlation function $g(r)$ for a typical amorphous structure. In
this figure we also include the pair correlation function
associated to those particles with icosahedral symmetry. Note that
the amplitude of the first maximum of the pair correlation
function calculated with the icosahedra is about three times
larger than the amplitude of the first maximum of $g(r)$
calculated with all particles. In addition, the position of the
first maximum for the icosahedral pair correlation function is
slightly smaller than the corresponding to all particles,
indicating that the icosahedral bonds are under compression.

The structural features found here (small clusters and chains of
icosahedra and the behavior of pair correlation functions), agree
with experimental findings in supercooled colloidal systems
undergoing NG \cite{WeitzScience}, \cite{Gasser}, suggesting that
icosahedra could also play a role in the amorphization of those
systems.

The structural characterization of amorphous matter has been one
of the major problems in condensed matter physics during the last
fifty years. In this sense, one of the best geometrical
descriptions is based in aggregates of polytetrahedral clusters
which have been invoked in order to describe the structure of
liquids, glasses and quasicrystals.

The geometrical features of the tetrahedra can be obtained through
Voronoi diagrams (Fig. 2d). To test for the regularity of
tetrahedral configurations in the amorphous phase we calculate the
$q_3$ order parameter, which is sensitive of local tetrahedral
order around the tetrahedron's center ($q_3=\sqrt{5}/3$ for a
regular tetrahedron). The inset of Fig. 4 shows the
$q_3$-distribution for our system. The sharp peak of the
distribution at $q_{3} \sim 0.73$ clearly indicates that regular
tetrahedra are the fundamental units of the amorphous states
obtained via SD. The analysis of the tetrahedral configurations
also shows that most of the particles ($\sim 90 \%$) are involved
in at least one regular tetrahedron, indicating the presence of an
amorphous phase characterized by a dense polytetrahedral structure
\cite{Anikeenko}.

The inset of Fig. 4 also shows a pair correlation function
calculated through the centres of tetrahedra. After a few
oscillations $g(r)$ rapidly decays towards the asymptotic value
for $r\gtrsim a$, showing that the tetrahedra aggregates in local
configurations without long range order. By using real space plots
we observe configurations of tetrahedra packed on common faces,
forming disordered clusters (Fig. 2d). Similarly to dense
disordered packing of hard spheres, here we also found that the
aggregates are mainly formed by five-fold rings obtained by
packing five tetrahedra around a common edge \cite{Anikeenko}.

In conclusion, we have presented evidence of a new mechanism of
crystallization during SD, induced by amorphous precursors. Local
icosahedral clusters and amorphous precursors are kinetically
favored by the early dynamics of SD. Crystalline particles
aggregate on these precursors in a similar way as heterogeneous
nucleation on spherical colloidal seeds. Our results are in
qualitative agreement with recent NG work in colloidal systems
\cite{TanakaPNAS}. Although the relaxational mechanism studied
here can be easily confused with conventional NG, it is hoped that
the  phenomenon of crystallization in the neighborhood of the
spinodal can be experimentally observed in systems with a slow
dynamics and a high length-scale selectivity like hard-sphere
colloidal suspension or block copolymers.

We thank V. Vitelli for helpful discussions, and support by the
Universidad Nacional del Sur (UNS), the National Research Council
of Argentina (CONICET), the National Agency of Scientific and
Technique Promotion (ANPCyT), and the FOM-Shell Industrial
Partnership Programme.


\begin{thebibliography}{9}
\bibitem{Comment} Here crystalline particles are identified by using an algorithm that finds ordered
regions independently of the symmetry \cite{WeitzScience},
\cite{FrenkelPRL}.
\end{thebibliography}
\end{document}